# Server Consolidation: An Approach to Make Data Centers Energy Efficient & Green

Mueen Uddin, Azizah Abdul Rahman

**Abstract -** Data centers are the building blocks of IT business organizations providing the capabilities of centralized repository for storage, management, networking and dissemination of data. With the rapid increase in the capacity and size of data centers, there is a continuous increase in the demand for energy consumption. These data centers not only consume a tremendous amount of energy but are riddled with IT inefficiencies. All data center are plagued with thousands of servers as major components. These servers consume huge energy without performing useful work. In an average server environment, 30% of the servers are "dead" only consuming energy, without being properly utilized. Their utilization ratio is only 5 to 10 percent. This paper focuses on the use of an emerging technology called virtualization to achieve energy efficient data centers by providing a solution called server consolidation. It increases the utilization ratio up to 50% saving huge amount of energy. Server consolidation helps in implementing green data centers to ensure that IT infrastructure contributes as little as possible to the emission of green house gases, and helps to regain power and cooling capacity, recapture resilience and dramatically reducing energy costs and total cost of ownership.

—  — —— — — — — —  ◆  — — — — — — — — — —

## 1. INTRODUCTION

Data centers are the building blocks of any IT business organization, providing capabilities of centralized storage, backups, management, networking and dissemination of data in which the mechanical, lighting, electrical and computing systems are designed for maximum energy efficiency and minimum environmental impact [1]. Data centers are found in nearly every sector of the economy, ranging from financial services, media, high-tech, universities, government institutions, and many others. They use and operate data centers to aid business processes, information management and communication functions [2]. Due to rapid growth in the size of the data centers there is a continuous increase in the demand for both the physical infrastructure and IT equipments, resulting in continuous increase in energy consumption.

Data center IT equipment consists of many individual devices like Storage devices, Servers, chillers, generators, cooling towers and many more. But Servers are the main consumers of energy because they are in huge number and their size continuously increases with the increase in the size of data centers.

As new servers are being added continuously into data centers without considering the proper utilization of already installed servers, it will cause an unwanted and unavoidable increase in the energy consumption, as well as increase in physical infrastructure like over-sizing of heating and cooling equipments. This increased consumption of energy causes an increase in the production of green house gases which are hazardous for environmental health. Hence it not only consumes space, energy, but also cost environmental stewardship.

Virtualization technology is now becoming an important advancement in IT especially for business organizations and has become a top to bottom overhaul of the computing industry. Virtualization combines or divides the computing resources of a server based environment to provide different operating environments using different methodologies and techniques like hardware and software partitioning or aggregation, partial or complete machine simulation, emulation and time sharing [3].

It enables running two or more operating systems simultaneously on a single machine. Virtual machine monitor (VMM) or hypervisor is a software that provides platform to host multiple operating systems running concurrently and sharing different resources among each other to provide services to the end users depending on the service levels defined before the processes [4].

Virtualization and server consolidation techniques are proposed to increase the utilization of underutilized servers so as to decrease the energy consumption by data centers and hence reducing the carbon footprints [4].

Section 2 provides a detailed background of the problem and emphasizes the need for implementing virtualization technology to save energy and cost. Section 3 describes the solution of the problem and proposes a methodology of categorizing the resources of data center into different resource pools, and analysis of the results to prove the benefits of server consolidation. Section 4 describes the process of implementing virtualization technology in a data center. In the last conclusions and recommendations are given.

## 2. LITERATURE REVIEW

In recent years the commercial, organizational and political landscape has changed fundamentally for data centre operators due to a confluence of apparently incompatible demands and constraints.

The energy use and environmental impact of data centers has recently become a significant issue for both operators and policy makers. Global warming forecasts that rising temperatures, melting ice and population dislocations due to the accumulation of greenhouse gases in our atmosphere from use of carbon-based energy. Unfortunately, data centers represent a relatively easy target due to the very high density of energy consumption and ease of measurement in comparison to other, possibly more significant areas of IT energy use. Policy makers have identified IT and specifically data centre energy use as one of the fastest rising sectors. At the same time the commodity price of energy has risen faster than many expectations. This rapid rise in energy cost has substantially impacted the business models for many data centers. Energy security and availability is also becoming an issue for data centre operators as the combined pressures of fossil fuel availability, generation and distribution infrastructure capacity and environmental energy policy make prediction of energy availability and cost difficult [5].

As corporations look to become more energy efficient, they are examining their operations more closely. Data centers are found a major culprit in consuming a lot of energy in their overall operations. In order to handle the sheer magnitude of today's data, data centers have grown themselves significantly by continuous addition of thousands of servers. These servers are consuming much more power, and have become larger, denser, hotter, and significantly more costly to operate [6]. An EPA Report to Congress on Server and Data Center Energy Efficiency completed in 2007 estimates that data centers in USA consume 1.5 percent of the total USA electricity consumption for a cost of $4.5 billion [7]. From the year 2000 to 2006, data center electricity consumption has doubled in the USA and is currently on a pace to double again by 2011 to more than 100 billion kWh, equal to $7.4 billion in annual electricity costs [8].

In USA the number of server farms (data centers) has increased from 7000 to 10000 and they are increasing as the demand from the end users increases. Gartner group emphasizes on the rising cost of energy by pointing out that, there is a continuous increase in IT budget from 10% to over 50% in the next few years. Energy increase will be doubled in next two years in data centers [9]. The statistics Cleary shows that the yearly cost of power and cooling bill for servers in data centers are around $14billion and if this trend persists, it will rise to $50billion by the end of decade [10]. Between the years 2000 and 2006, the number of servers grew from 5.5 million to 10.9 million [11].

With the increase in infrastructure and IT equipment, there is a considerable increase in the energy consumption by the data centers, and this energy consumption is doubling every five years. [12]. Data centers use nearly 10 to 30 times more energy per square foot than office space [13]. Today's data centers are big consumer of energy and are filled with high density, power hungry equipment. If data center managers remain unaware of these energy problems then the energy costs will be doubled between 2005 and 2011. If these costs continue to double every five years, then data center energy costs will increase to 1600 % between 2005 and 2025 [14]. Currently USA and Europe have largest data center power usage but Asia pacific region is rapidly catching up. [15].

### 2.1 Problem Statement

As some of the developing countries are facing huge energy crisis, the energy consumed by data centers have a great effect on their overall production of energy. This huge consumption of energy by data centers not only shortens the supply of power energy to other businesses but also contributes towards the shortage for data centers themselves. This energy consumption also contributes towards waste of energy and environmental stewardship.

There is a need to design a strategy that provides a solution to decrease the continuous demand and consumption of energy by data centers.

This paper proposes a new technique that combines the workload of multiple servers onto fewer servers by properly utilizing their efficiencies i.e. hardware and software efficiencies. The proposed strategy uses a new technology called server consolidation a type of virtualization to achieve energy efficiency and at the same time reducing the effect of green house gases making data centers greener.

## 3. PROPOSED WORK

Servers are the leading consumer of IT power in any data center. Data centers are plagued with thousands of the server's mostly underutilized, having utilization ratio of only 5 to 10%, consuming huge energy and generating huge amount of green house gases.

This paper focuses on the use of virtualization to overcome energy problems in data centers. It provides a mechanism to save huge amount of energy and at the same time increases the productivity of servers with little or no additional energy consumption.

The proposed strategy categorizes the servers into three resource pools that are innovation, production and mission critical servers depending on their workload and usage. After categorizing server consolidation is applied on all categories depending on their utilization ratio in the data center. This process reduces the number of servers by consolidating the load of multiple servers on one server. Finally a comparison of energy consumed by different servers is analyzed by a tool called SPECpower_ssj [16]. This software is used as a performance evaluation benchmarking standard for calculating energy efficiencies in data centers.

### 3.1 Server Categorization
The physical and IT equipments in data centers consume a lot of energy. All data centers are predominantly occupied by low cost underutilized volume servers also called x-86 servers. These servers play an important role in achieving and providing data center services to the end users and on the other hand consume a lot of energy. With the recent development and growth in the data centers their number continuously increases as the demand for storage, speed, backups and recovery increases.

In this paper a new technique is proposed for underutilized volume servers to categorize them on the basis of their workloads they perform and applications they execute. Although these classifications may vary from business to business, because servers are generally used to create, maintain and execute solutions on the behalf of businesses, architectures, processes and infrastructures.

### 3.1.1 Innovations Servers
These servers are mostly deployed in shared computing environments, where there is a provision of addition of new servers. The applications these servers execute require speed and flexibility and should be completed quickly. Innovation servers are deployed at locations where there is huge potential of inventing new products, modifying existing products, develop and enhance processes which are more competitive and productive. This creates space for addition of new servers hence their number tends to increase rapidly, without considering proper utilization of existing servers, consuming a lot of energy and producing huge amount of green house gases. These servers typically are referred to as test, deployment, quality assurance and volume servers. These servers are the mostly available servers in a data center.

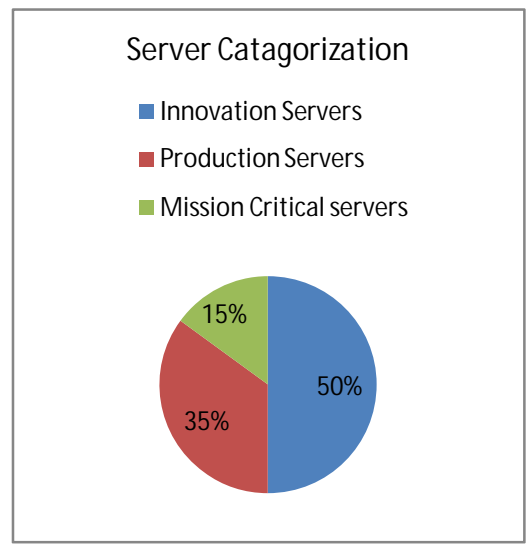

Figure1 Server Categorization

### 3.1.2 Production Servers
These are much more controlled and scalable servers deployed at locations where there is a less chance of addition of new servers. The Service level requirements for different applications and servers are more important than speed and flexibility. That's why their number is lesser than innovation servers.

### 3.1.3 Mission Critical Servers

Mission critical servers are the most powerful servers in any data center consuming more energy and power as compare to other types of servers. These servers handle critical real time jobs and applications where there is a chance of loss of memory and customer loyalty. These servers are normally small in number and have significant impact on overall business requirements.

### 3.2 Server Consolidation

After categorization of volume servers into different resource pools, the next step is server consolidation. The process of server consolidation always begins from innovation servers because these servers are mostly underutilized and remain idle for long durations of time. The other most important reason for applying server consolidation on innovation servers is that they are in huge number and require fewest computing resources.

In this paper, we analyzed a data center consisting of total 500 hundred servers. These servers are categorized according to their workloads and applications they execute.

After applying the proposed technique of server consolidation, results were generated by comparing the pre and post consolidation ratios of these servers.

Table 1 shows the utilization ratio and energy consumed in watts by each category of server before applying server consolidation.

Table1 Pre-Consolidation Ratio of servers

| Categories (Servers) | Innovation | Production | Mission Critical | Total energy |
|---|---|---|---|---|
| Server count | 250 | 175 | 75 | 500 |
| Utilization | 3% | 6% | 10% | 5% |
| Watts per server=173* server count | 43250 | 30275 | 12975 | 86500 |

Table 2 Post-Consolidation Ratio of servers

| Categories | Innovation | Production | Mission Critical | Total energy |
|---|---|---|---|---|
| Consolidation ratio | 15:1 | 10:1 | 5:1 | 10:1 |
| Post consolidation utilization | 50% | 50% | 50% | 50% |
| Post consolidation energy in watts | 3910 | 4140 | 3450 | 11500 |
| Energy saving | 39340 | 26135 | 9525 | 75000 |

Table 2 shows the utilization ratio and energy consumed in watts by each category of server after applying server consolidation.

### 3.3 Analysis of Results

There is no linear relationship between energy consumption and productivity output of a server. If a server is running at 5% utilization and target utilization rate is 50%, then energy consumption can be greatly reduced to almost 50%, by improving server utilization and turning off unused idle servers.

In table 1 it is shown that with average 5% utilization, every server consumes 173 watts of power; hence the total energy consumed by all 500 servers is 86500 watts. After applying server consolidation with consolidation ratios of 15:1 for innovation servers, 10:1 for production servers and 5:1 for mission critical servers, it is observed that there is an increase of 57 watts of more power, a total of 230 watts per server as compared to 173 watts before server consolidation with utilization ratio of 50%.

So total energy consumed by all servers after consolidation is only 11500 watts compared to 86500 watts before applying server consolidation, with a total saving of 75000 watts of energy. Further more if this utilization ratio is increased to 100%, then only 275 watts of energy will be consumed by every server thus an increase of only 102 watts of power with utilization ratio up to 100%.

SPECpower_ssj is a performance evaluation benchmarking software used for calculating energy efficiencies in data centers.

By applying SPECpower_ssj on the above results, it becomes very much clear that when servers are consolidated to an average ratio of 10:1, then there is a considerable reduction in the energy consumption from 1730 watts that is 10 servers with 173 watts per server to only 230 watts with only one server having utilization ratio of 50%. Hence server consolidation yields a saving of almost 1500 watts i.e. 87% of total energy.

$$1730-230=1500 \text{ watts}$$

Thus it is concluded that power utilization ratio is very less as compare to processor utilization ratio. Decreasing physical servers not only reduces energy consumption but also has great impact on the overall data center heating requirements, cooling load, as well as increases UPS backup time. It also increases performance efficiency, longer generator backup times, and reduces IT configuration of different interconnecting devices.

The other main advantage of reducing the power energy consumption is that, it reduces the emission of green house gases, which currently have become an alarming sign for the whole world because data centers are the main culprit of generating around 2% of the world's $CO_2$.

The implementation of Server consolidation does not require any budgeting or finance approval, platform and vendor selection, provisioning, installation, patching, security and networking.

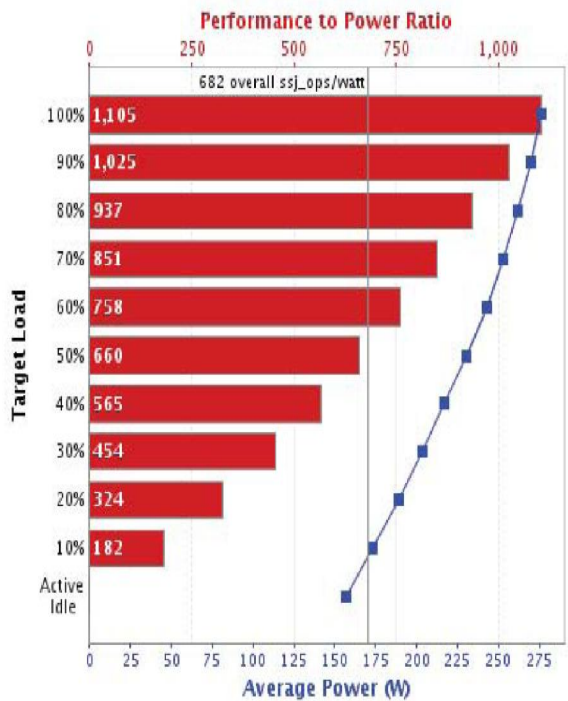

Figure 2 Performance to Power ratio

In the context of server virtualization, a study from VMware Information warehouse revealed that more than 20% of servers assessed by VMware capacity planner tool in 2007 were running below 0.5% utilization, and approximately 75% of servers assessed were running below 5% utilization [4]. At the same time huge number of servers remained idle for most of the time consuming huge energy without providing any processing.

These statistics highlight the importance of virtualization specially server consolidation, which is the most popular type of virtualization, because it creates an ample opportunity to reduce the number of physical servers, saving huge amount of energy and reducing carbon footprints.

This simplifies IT infrastructure and saves a lot of cost being wasted on providing power, heating and cooling. It also fastens the overall business time and services and provides a means for dynamic allocation of new capacity within minutes of a business service request.

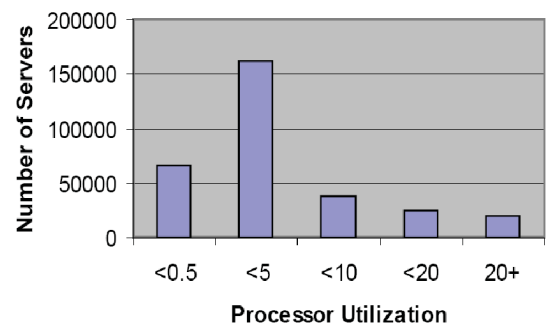

Figure 3 VMware Processor utilization

It is important to note that properly implementing virtualization technology at different levels in the data center betters the service levels needed for business perspective.

## 4. PROCESS OF IMPLEMENTING SERVER CONSOLIDATION

The last part of this paper discusses the importance of virtualization and the process of implementing it in a business firm. The process involves many prerequisites need to be fulfilled.

### 4.1 Identify Server Population
The process of virtualization starts by creating an inventory of all servers, resources they require, available resources and their associated workloads, this process is called discovery process. The inventory process includes both utilized and idle servers. It also includes information related to

- Make and Model of the Processor

- Types of processors (socket, Core, Threads, Cache)
- Memory size and speed
- Network type (Number of ports, speed of each port)
- Local storage (number of disk drives, capacity, RAID)
- Operating system and their patch levels (service levels)
- Applications installed
- Running services

Different tools are available to automate the inventory and discovery process. Some of them are:

- Microsoft baseline security analyzer (MBSA)
- Microsoft assessment and planning toolkit (MAP)
- VMware guided consolidation (VCP)
- CIRBA's Power Recon and plate spin's.

### 4.2 Categorizing Server Resources

After creating server inventory information, the next step is to categorize the servers and their associated resources and workloads into resource pools. This process is performed to avoid any technical political, security, privacy and regulatory concern between servers, which prevent them from sharing resources. Some of the server categories are:

- Network Infrastructure servers
- Terminal servers
- File and Print servers
- Application servers
- Web servers
- Ideality management servers
- Collaboration servers
- Database servers

### 4.3 Categorizing Application resources

After categorizing servers into different resource pools, applications will also be categorized as:
- Commercial versus in-house
- Custom applications
- Legacy versus updated applications
- Infrastructure applications
- Support to business applications
- Line of business applications
- Mission critical applications

### 4.4 Allocation of Computing Resources

After creating the workloads, the next process is to allocate computing resources required by these different workloads and then arranging them in normalized form, but for normalization the processor utilization should be at least 50%.

It is very important to normalize workloads so as to achieve maximum efficiency in terms of energy, cost and utilization [4].

The formula proposed in this paper for normalization is to multiply utilization ratio of each server by total processor capacity that is (maximum processor efficiency * number of processors * number of cores).

Once all of the servers are grouped into appropriate categories and sub-categories, then next step will be to assign utilization objectives. The utilization targets may be set at or above 50% for the Innovation pool of servers, 25% to 50% for the Production pool of servers while 25% to 30% for mission critical servers.

The consolidation ratio will be driven by the utilization percentages for not just processors but for the other components in the servers as well.

It's important to note that consolidation rates are much higher for innovation servers thus saving a lot of energy. The Innovation environment normally has the most servers, the lowest utilization, and the highest tolerance for error. As data center costs rise regarding energy consumption and availability of energy, the numbers of virtualized servers may increase to allow more processing at lower levels of energy consumption.

## 5. CONCLUSION

In this paper we proposed a new technique to categorize servers and their associated resources by using a new formula. Then we implemented the proposed technique over a data center and analyzed the results using SPECpower_ssj. The results clearly show a saving of almost 75000 watts and utilization ratio of 50%.

So it is concluded that by properly implementing server consolidation, not only a huge amount of energy is saved but the effect of green house gases is also reduced, which makes data center more efficient, greener and cost effective

Server consolidation not only needs to characterize the workloads that are planned to be consolidated, but also target the environments into which the workloads are to be applied. It is important to determine the type of servers, their current status whether idle or busy, how much it will cost to implement server consolidation, the type of technology needed to achieve the service levels required and finally meet the security/privacy objectives. It is also important for the data center to check whether it has the necessary infrastructure to handle the increased power and cooling densities arise due to the implementation of virtualization.

It is also important to consider the failure of single consolidated server, because it is handling the workload of multiple applications.

Virtualization poses many challenges to the data center physical infrastructure like dynamic high density, under-loading of power/cooling systems, and the need for real-time rack-level management. These challenges can be met by row-based cooling, scalable power and predictive management tools. These solutions are based on design principles that simultaneously resolve functional challenges and increase efficiency.

## REFERENCES

[1] Data Center Journal (n.d.), "What is a data center", Data Center Journal, available at: http://datacenterjournal.com/index.php?option=com_content&task=view&id=63&Itemid= 147 (accessed March 16, 2009).

[2] T. Daim, J. Justice, M. Krampits, M. Letts, G. Subramanian, M. Thirumalai, "Data center metrics An energy efficiency model for information technologu managers", Management of Environmental Quuality, Vol.20 No.6, 2009.

[3] Amit singh,, "An introduction to virtualization",http://www.kernelthread.com/publications/virtualization, 2004.

[4] Green Grid, "Using virtualization to improve data center efficiency", available at: www.thegreengrid.org/ (accessed February 2009).

[5] L. Newcombe, "Data centre energy efficiency metrics", 2009.

[6] W. McNamara, G. Seimetz, K. A. Vales, "Best Practices for Creating The Green Data Center", 2008.

[7] EPA Report to Congress on Server and Data Center Energy Efficiency – Public Law109-431, Environmental Protection Agency, Washington, DC, 2007.

[8] Tung, T. Data Center Energy Forecast, Silicon Valley Leadership Group, San Jose, CA, 2008.

[9] Gartner, "Cost Optimization and Beyond: Enabling Business Change and the Path to Growth", A Gartner Briefing, Gartner, London, 2009.

[10] EPA Report to Congress on Server and Data Center Energy Efficiency – Public Law109-431, Environmental Protection Agency, Washington, DC, 2009.

[11] Koomey, J.G. Estimating Total Power Consumption by Servers in the US and the World,Lawrence Berkeley National Laboratory, Berkeley, CA, 2007.

[12] Gartner, "Sustainable IT", A Gartner Briefing, Gartner, Dublin, 2008.

[13] Caldow, J. "The greening of Government: a study of how governments define the green agenda", p. 8, available at:www01.ibm.com/industries/government/ieg/pdf/green_gov_agenda.pdf, 2008.

[14] Kumar, R. Media Relations, Gartner, available at:www.gartner.com/it/page.jsp?id¼781012, 2008.

[15] Fehrenbacher, K. STRUCTURE 08: Data Center Power Guru Jonathan Koomey,gigaom.com, available at:http://gigaom.com/2008/06/25/structure-08-data-center-powerguru-jonathan-koomey, 2008.

[16] SPEC SSJ Design, SPEC.org, available at:www.spec.org/power_ssj2008/docs/SPECpower_ssj2008-Design_ssj.pdf, 2008.